\documentclass{article}

\usepackage{authblk}
\usepackage{amsmath,amsfonts,amssymb}
\usepackage{setspace}
\usepackage[margin=1in]{geometry}
\usepackage{graphicx}
\usepackage{xcolor}
\usepackage{chngcntr}
\usepackage{subcaption}
\usepackage{float}
\usepackage{hyperref}

\usepackage{multibib}

\title{A Milestone-Based Framework for Characterizing Time-Varying Treatment Effects in Immunotherapy Trials}

\author[1,4]{Yi-Cheng Tai, PhD}
\author[2]{Weijing Wang, PhD$^{*}$}
\author[3]{Jedd D. Wolchok, MD PhD}
\author[1]{Martin T. Wells, PhD$^{*}$}

\affil[1]{Department of Statistics and Data Science, Cornell University, New York, USA}
\affil[2]{Institute of Statistics, National Yang Ming Chiao Tung University, Hsinchu City, Taiwan}
\affil[3]{Sandra and Edward Meyer Cancer Center, Weill Cornell Medicine, New York, USA}
\affil[4]{Department of Statistics, National Chengchi University, Taipei, Taiwan}

\date{}

\protect \begin{document}

\onehalfspacing
\maketitle

\noindent $^{*}$Corresponding authors: Martin T. Wells, mtw1@cornell.edu; Weijing Wang, wjwang@stat.nycu.edu.tw

\begin{abstract}
Immune checkpoint inhibitor--based therapies often produce heterogeneous survival responses, including early risk, delayed treatment benefit, and durable long-term survival in a subset of patients. In these settings, conventional summary measures such as the hazard ratio may not adequately describe how treatment effects evolve over follow-up. We propose a milestone-based framework that separates long-term survival beyond a clinically meaningful time point from earlier outcomes and provides a practical way to characterize patient heterogeneity in treatment response. The framework summarizes treatment differences through milestone survival probabilities and, among patients who do not reach the milestone, characterizes short-term treatment ordering over time using a tau-based summary that helps identify hazard reversal. We illustrate the approach using reconstructed individual-level data from three landmark phase III trials: CheckMate~067, CheckMate~227, and CLEAR. Across these examples, the framework captures patterns that are difficult to summarize with conventional measures, including settings in which early disadvantage coexists with later durable benefit. It also helps clarify when treatment benefit begins to emerge and how short-term and long-term effects differ within the same trial. This approach provides a clinically interpretable and statistically principled way to evaluate heterogeneous and time-varying treatment effects in oncology trials with nonproportional hazards.
\end{abstract}


\section*{Translational Relevance}

Immunotherapy trials often exhibit heterogeneous treatment responses and nonproportional hazards, with early risk and delayed long-term benefit occurring within the same study. Conventional summary measures such as hazard ratios do not distinguish these patterns clearly and may therefore obscure clinically relevant treatment effects. The milestone-based framework developed in this study separates durable survival beyond a clinically meaningful time point from earlier outcomes. Among patients who do not reach the milestone, it further characterizes how treatment effects evolve over follow-up. Applied to landmark phase III trials, this approach clarifies benefit--risk trade-offs, identifies when treatment benefit begins to emerge, and provides a more interpretable description of short-term and long-term treatment effects within the same trial. In this way, the framework may improve communication of trial results and support clinical interpretation in modern oncology studies.
\section{Introduction}

\subsection{Background}

Interpretation of treatment effects in modern oncology trials has become increasingly challenging, particularly in studies of immune checkpoint inhibitor--based therapies. Kaplan--Meier (KM) curves in these trials may overlap or cross early, separate only after a delay, and sometimes exhibit a plateau in the tail consistent with durable survival in a subset of patients. These features suggest that treatment effects may change over time and that patient responses are heterogeneous. \cite{quinn2020current, oquigley2022testing, THELANCETONCOLOGY2024147}
Conventional survival methods are often inadequate in this setting. When the proportional hazards (PH) assumption is violated, the hazard ratio (HR) may not provide a meaningful summary because it averages over qualitatively different treatment phases. In immunotherapy trials, treatment effects may include early disadvantage, delayed benefit, or durable long-term survival in only a subset of patients. As a result, a single summary measure can obscure clinically important differences between short-term risk and later benefit. The log-rank test may also lose power under nonproportional hazards. \cite{oquigley2022testing, uno2014moving,stensrud2019limitations,bartlett2020hazards}

These challenges have motivated alternative approaches for survival analysis under nonproportional hazards. Restricted mean survival time (RMST) has become increasingly popular because it provides an interpretable treatment-effect summary based on the difference in mean survival time up to a prespecified horizon, consistent with the ICH estimand framework. \cite{Royston2011RestrictedMean,EMAE92020} Other work has focused on hypothesis testing in survival settings that depart from the proportional hazards framework. These methods are designed to retain good performance under a range of nonproportional hazard alternatives. \cite{oquigley2022testing,Lin2020,Roychoudhury2021,Mukhopadhyay2022}
As an alternative perspective, Tai et al.\ proposed the tau process \cite{tai2023twosample}, motivated by its connection to the Wilcoxon--Mann--Whitney statistic and Kendall's tau. \cite{gibbons2014nonparametric} The tau process summarizes, over time, the difference in pairwise ordering probabilities between treatment groups. Its temporal slope reflects the local hazard difference and can therefore help identify the timing of hazard reversal. In clinical terms, this provides a dynamic description of when an experimental treatment begins to show advantage relative to control. The tau process may be viewed as a time-evolving generalization of net benefit, extending beyond a single summary value to describe how comparative treatment benefit unfolds during follow-up. \cite{pocock2012win,martinussen2025debiased}

In practice, clinicians and investigators are interested in long-term benefit, but the meaning of ``long-term'' requires a more explicit definition. Beyond overall survival differences, the key question is whether a treatment increases the probability of surviving beyond a clinically meaningful time point and how outcomes differ among patients who do not. This issue is especially relevant in immunotherapy trials, where some patients derive sustained benefit whereas others experience progression or death relatively early.
Several methods have been proposed to evaluate long-term survival patterns. RMST-based analyses have been adapted to settings with long-term survivors through clinically relevant time windows. \cite{Horiguchi2018Quantification,horiguchi2023assessing,Kloecker2020Uses} Because RMST is time-integrated, survival differences accumulate over follow-up. In settings with long-term survivors, a persistent tail difference may therefore lead to a larger RMST contrast as the time window increases. Cure-mixture approaches provide a natural framework for separating long-term from short-term treatment effects. Within this framework, Cox--TEL compares tail probabilities for long-term survivors and uses a short-term hazard ratio for the susceptible subgroup under a proportional hazards assumption. \cite{hsu2021development,lin2020changing,lin2022cox} The tau approach has also been extended to this setting, offering a model-free alternative for the susceptible subgroup that does not require a proportional hazards assumption. \cite{tai2024estimandbasedinference}
In practice, however, identification of the long-term component usually requires sufficiently mature follow-up; when follow-up is limited, additional assumptions or extrapolation methods may be needed. \cite{EscobarBach2019}

To address this gap, we introduce a milestone-based framework that uses clinically meaningful time points to distinguish long-term survival from earlier outcomes. Rather than relying on a cured subgroup that cannot be directly identified from observed follow-up, the framework classifies patients by whether they survive beyond a prespecified milestone and then evaluates both the long-term survivor fraction and treatment effects among those who do not. This provides a clinically interpretable way to characterize heterogeneous and time-varying treatment effects in immunotherapy trials.

\subsection{Patient Heterogeneity in Extended Follow-Up}
\label{subsec:patient_heterogeneity}

Extended follow-up is essential for understanding whether treatment benefit is durable, delayed, or attenuated over time. This is particularly important in immunotherapy trials, where treatment effects may emerge late or change substantially over follow-up. The CheckMate 067 trial in advanced melanoma illustrates this clearly, with major follow-up reports at 4 years, 5 years, 6.5 years, and 10 years. \cite{Drye2014Transitioning,hodi2018nivolumab,Larkin2019,Wolchok2022,Wolchok2024} Such extended observation provides an opportunity to examine how survival patterns evolve and whether apparent long-term benefit remains stable with increasing data maturity.

The cure-mixture model provides a conceptual framework for separating patients into more homogeneous groups, typically described as cured and susceptible subpopulations. In this setting, ``cure'' refers to a subgroup that would never experience the event of interest under sufficiently long follow-up. In practice, however, this distinction is difficult to verify. Although formal procedures for assessing follow-up adequacy have been proposed, their assumptions are often restrictive. \cite{MallerZhou1994} As Farewell emphasized, long-tailed survival curves make it difficult to distinguish truly cured patients from censored patients with long event-free follow-up, leading to an inherent non-identifiability problem. \cite{farewell1986mixture}
The milestone-based perspective offers a practical and clinically interpretable alternative. By defining long-term survival relative to a clinically specified time point, it distinguishes durable benefit from earlier outcomes without requiring a literal cure interpretation. Patients who survive beyond the milestone are regarded as long-term survivors beyond that time point, whereas those who experience the event before the milestone are treated as the short-term or susceptible subgroup. This framework also allows reanalysis across different follow-up durations, providing additional insight into treatment heterogeneity, delayed benefit, and the trade-off between early risk and durable survival.

In the analyses that follow, we apply this framework to landmark phase III oncology trials with distinct survival patterns, including early overlap, curve crossing, and late plateau formation. Our goal is not only to quantify long-term survival beyond clinically meaningful milestones, but also to clarify treatment-effect patterns that conventional summaries may not capture. The milestone framework is intended as an interpretive tool for understanding heterogeneous and time-varying treatment effects in modern oncology trials.

\section{Materials and Methods}

\subsection{Milestone-Based Mixture Framework}

We propose a milestone-based framework to separate long-term survival from earlier outcomes using a clinically prespecified time point \(m\). Patients who remain event-free beyond \(m\) are classified as long-term survivors beyond the milestone, whereas patients who experience the event by \(m\) are classified as the short-term or susceptible subgroup. This perspective is motivated by the clinical importance of distinguishing durable benefit from earlier risk patterns in oncology trials. \cite{Chen2015MilestoneSurvival,10.1001/jamaoncol.2015.4345}

In the one-sample setting, the survival function of \(T\) can be decomposed as
\[
S(t)=\Pr(T>t \mid T\le m)\Pr(T\le m)+\Pr(T>t \mid T>m)\Pr(T>m).
\]
Because \(\Pr(T>t \mid T>m)=1\) for \(t\le m\), this becomes
\begin{equation}
\label{KM_deomposition}
S(t)=S_a(t;m)\{1-\eta(m)\}+\eta(m), \qquad t\le m,
\end{equation}
where
\[
S_a(t;m)=\Pr(T>t \mid T\le m)
\]
denotes the conditional survival function among patients who do not reach the milestone, and
\[
\eta(m)=\Pr(T>m)
\]
denotes the proportion of long-term survivors beyond the milestone.

This decomposition extends naturally to a two-sample setting indexed by \(j=0,1\). Let
\[
S_j(t)=\Pr(T_j>t)
\]
denote the survival function in group \(j\). Then, for \(t\le m\),
\begin{equation}
\label{KM_deomposition_two_sample}
S_j(t)=S_{a,j}(t;m)\{1-\eta_j(m)\}+\eta_j(m), \qquad j=0,1,
\end{equation}
where $\eta_j(m)=\Pr(T_j>m)$ and 
\[
S_{a,j}(t;m)=\Pr(T_j>t \mid T_j\le m).
\]
The framework extends directly to more than two groups, as in the trial applications presented later. Within this formulation, the contrast
\[
\eta_j(m)-\eta_0(m)
\]
quantifies the treatment effect on the proportion of patients surviving beyond the milestone. In contrast, comparisons between \(S_{a,j}(t;m)\) and \(S_{a,0}(t;m)\) describe treatment differences among patients who do not reach the milestone.

To characterize time-varying treatment effects, we extend the tau process proposed by Tai et al. \cite{tai2023twosample} to the milestone framework. The original tau process is a model-free estimand that is particularly useful when the proportional hazards assumption is inappropriate. It is defined as
\begin{equation}
\label{eq:tauprocess}
\tau(t)=\Pr(T_0<T_1\wedge t)-\Pr(T_1<T_0\wedge t)
=\int_0^t S_1(u)\,dF_0(u)-\int_0^t S_0(u)\,dF_1(u),
\end{equation}
where \(F_j(t)=1-S_j(t)\) denotes the cumulative distribution function in group \(j\).
A positive value of \(\tau(t)\) indicates that, up to time \(t\), patients in Group 1 are more likely to survive longer than patients in Group 0. Its derivative,
\[
\tau'(t)=S_0(t)S_1(t)\{\lambda_0(t)-\lambda_1(t)\},
\]
relates the local slope of the tau curve to the hazard difference between groups, so that a change in sign of \(\tau'(t)\) indicates hazard reversal. When long-term survivors are present, however, interpretation of both hazards and the overall tau process may be complicated by heterogeneity in the risk set, as discussed in Section~\ref{subsec:patient_heterogeneity}.
To address this issue, we define a short-term tau process within the milestone framework for patients who do not reach the milestone. Specifically, among the susceptible subgroup defined by \(\{T_0\le m,\,T_1\le m\}\), we define, for \(t<m\),
\begin{align}
\tau_a(t;m)
&=\Pr(T_1\wedge t>T_0 \mid T_0\le m, T_1\le m)
-\Pr(T_0\wedge t>T_1 \mid T_0\le m, T_1\le m) \nonumber\\
&=\int_0^t S_{a,1}(u;m)\,dF_{a,0}(u;m)
-\int_0^t S_{a,0}(u;m)\,dF_{a,1}(u;m),
\end{align}
where
\[
F_{a,j}(t;m)=1-S_{a,j}(t;m).
\]
A positive value of \(\tau_a(t;m)\) indicates that, among patients who do not reach the milestone, Group 1 tends to have longer event times than Group 0 up to time \(t\). Within the milestone framework, treatment effects are thus summarized by two complementary components: the long-term contrast \(\eta_1(m)-\eta_0(m)\) and the short-term trajectory \(\tau_a(t;m)\).
The local slope of the short-term tau process is
\begin{equation}
\label{tau_slope}
\tau_a'(t;m)=S_{a,0}(t;m)S_{a,1}(t;m)\{\lambda_{a,0}(t;m)-\lambda_{a,1}(t;m)\},
\end{equation}
where \(\lambda_{a,j}(t;m)\) is the hazard function for \(T_j\) conditional on \(T_j\le m\), given by
\begin{equation}
\label{hazard_group}
\lambda_{a,j}(t;m)=\frac{f_{a,j}(t;m)}{\Pr(T_j\in[t,m])},
\end{equation}
for $t<m$ and $j=0,1$
with
\[
f_{a,j}(t;m)=\frac{f_j(t)}{\Pr(T_j\le m)},
\]
and \(f_j(t)\) is the density function of \(T_j\).
A positive slope of \(\tau_a(t;m)\) indicates that, within the susceptible subgroup, Group 1 has a lower hazard than Group 0 at that time. Accordingly, a change in sign of \(\tau_a'(t;m)\) marks hazard crossing within the milestone-defined short-term population. In this setting, delayed treatment effect corresponds to an early negative slope followed by a sustained positive slope, indicating that treatment benefit emerges after an initial period of disadvantage.

For estimation, let \(\hat S_j(t)\) denote the Kaplan--Meier estimator in group \(j\). The long-term survivor proportion is estimated by
\[
\hat\eta_j(m)=\hat S_j(m).
\]
Using the decomposition in (\ref{KM_deomposition_two_sample}), the conditional survival function among patients who do not reach the milestone is estimated by
\begin{equation}
\label{S_a_est}
\hat S_{a,j}(t;m)=\frac{\hat S_j(t)-\hat\eta_j(m)}{1-\hat\eta_j(m)}.
\end{equation}
The corresponding short-term tau process is estimated by
\begin{equation}
\label{tau_a_est}
\hat\tau_a(t;m)=\int_0^t \hat S_{a,1}(u;m)\,d\hat F_{a,0}(u;m)
-\int_0^t \hat S_{a,0}(u;m)\,d\hat F_{a,1}(u;m),
\end{equation}
where
\[
\hat F_{a,j}(t;m)=1-\hat S_{a,j}(t;m), \qquad j=0,1.
\]
In settings with three groups, we use the notation \(\hat\tau_{a,j}(t;m)\) for \(j=1,2\), with Group 0 as the reference. In our analyses, point estimates, confidence intervals, and \(P\) values are obtained from 10{,}000 bootstrap resamples using the \texttt{tauProcess} package. \cite{tauProcess}
Additional methodological details, including supplementary notation, estimation remarks, and comments on late follow-up interpretation, are provided in the Supplementary Materials.

\subsection{Treatment Assessment Across Evolving Milestones}

Because long-term treatment benefit often becomes clearer with extended follow-up, we also examine how interpretation changes as the milestone time \(m\) increases. Let
\[
\xi(m)=I(T\le m),
\]
where \(\xi(m)=0\) indicates survival beyond the milestone and \(\xi(m)=1\) indicates failure by the milestone. Then the study population can be partitioned as
\[
\Omega=\Omega_1^m\cup\Omega_0^m=\{\xi(m)=1\}\cup\{\xi(m)=0\}.
\]
If \(m_a<m_b\), then
\[
\{\xi(m_a)=0\}\supseteq \{\xi(m_b)=0\},
\]
so that the set of patients classified as long-term survivors cannot increase as the milestone is extended. Equivalently,
\[
\eta(m_a)\ge \eta(m_b).
\]
In practical terms, this means that additional follow-up may reveal later events and reclassify some patients from the long-term subgroup into the susceptible subgroup.

Within the two-group setting, a positive value of
\[
\eta_1(m)-\eta_0(m)
\]
indicates a higher proportion of long-term survivors in Group 1. Meanwhile, a positive \(\tau_a(t;m)\) indicates a short-term advantage among patients who do not reach the milestone. As \(m\) increases, the long-term contrast \(\eta_1(m)-\eta_0(m)\) depends on the tail behavior of the two groups rather than following a fixed pattern. If, at the same time, \(\tau_a(t;m)\) remains positive or increases over follow-up, this pattern may be consistent with longer-tail survival in Group 1 relative to Group 0, in the sense described by Farewell for prolonged survival within the susceptible subgroup. \cite{farewell1986mixture}

Interpretation of late follow-up also depends on data maturity. Del Paggio and Tannock emphasized that reliable interpretation of the tail of a survival curve requires sufficiently mature follow-up and an adequate number of events. \cite{DelPaggio2021CautionaryTails} Kaplan--Meier curves are commonly presented with numbers at risk for this reason. Good-practice guidance has suggested that approximately 10\% to 20\% of patients should remain at risk in the tail for these late-time estimates to be considered reasonably stable. \cite{DelPaggio2021CautionaryTails,10.1001/jamaoncol.2015.4345,Pocock2002}

\section{Results}
We applied the proposed framework to reconstructed individual-level data from three phase III trials with distinct survival patterns: CheckMate 067, CheckMate 227, and CLEAR. For CheckMate 067, we analyzed the 6.5-year follow-up for progression-free survival (PFS) and overall survival (OS), together with the final 10-year report for OS and melanoma-specific survival (MSS). \cite{Wolchok2022,Wolchok2024} For CheckMate 227, we focused on the 5-year OS analysis. \cite{Brahmer2023} For CLEAR, we analyzed both PFS and OS using the final prespecified OS report. \cite{Motzer2024} Additional analyses, including proportional hazards diagnostics, dRMST processes, and supplementary subgroup results, are provided in the Supplementary Materials.

\subsection{CheckMate 067}

\begin{figure}[H]
\centering
\begin{minipage}{.49\textwidth}
    \centering
    \includegraphics[width=\linewidth]{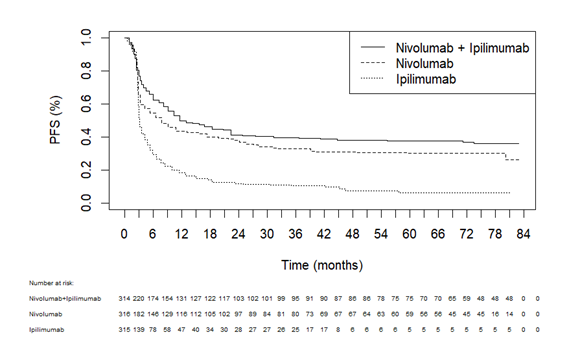}
    \subcaption{Digitized Progression-Free Survival Curves.}
    \label{fig:Trial_067_PFSdigitized}
\end{minipage}%
\hfill
\begin{minipage}{.49\textwidth}
    \centering
    \includegraphics[width=\linewidth]{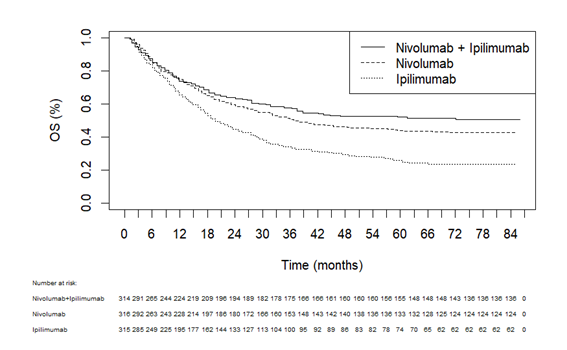}
    \subcaption{Digitized Overall Survival Curves.}
    \label{fig:Trial_067_OSdigitized}
\end{minipage}%

\vspace{1em} 

\begin{minipage}{.49\textwidth}
    \centering
    \includegraphics[width=\linewidth]{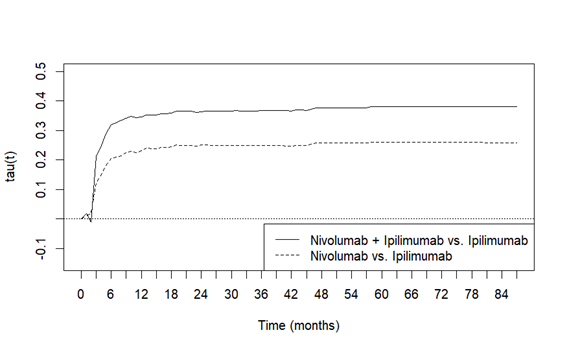}
    \subcaption{Tau Curves for PFS.}
    \label{fig:Trial_067_tau_PFS}
\end{minipage}%
\hfill
\begin{minipage}{.49\textwidth}
    \centering
    \includegraphics[width=\linewidth]{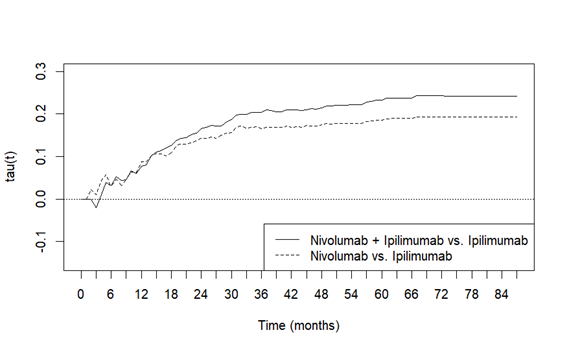}
    \subcaption{Tau Curves for OS.}
    \label{fig:Trial_067_tau_OS}
\end{minipage}

\vspace{1em} 

\begin{minipage}{.49\textwidth}
    \centering
    \includegraphics[width=\linewidth]{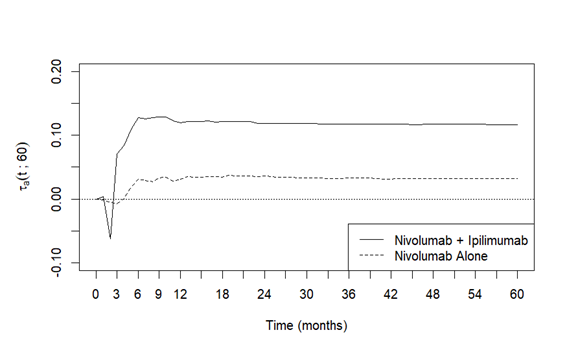}
    \subcaption{Tau Curve for PFS in Patients Not Reaching the 60-Month Milestone.}
    \label{fig:Trial_067_susceptible_tau_PFS}
\end{minipage}%
\hfill
\begin{minipage}{.49\textwidth}
    \centering
    \includegraphics[width=\linewidth]{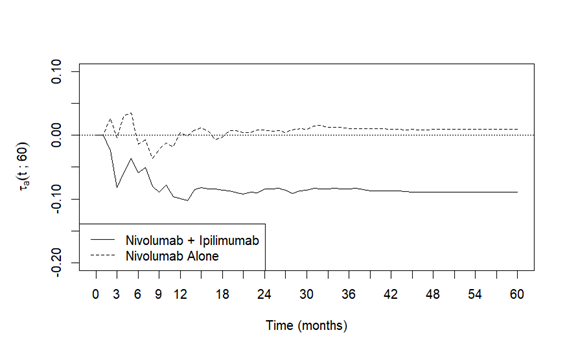}
    \subcaption{Tau Curve for OS in Patients Not Reaching the 60-Month Milestone.}
    \label{fig:Trial_067_susceptible_tau_OS}
\end{minipage}
\caption{Comprehensive analysis from the 6.5-year follow-up of the CheckMate 067 trial, presenting metrics based on OS and PFS, along with their tau curves. Data for OS are derived from Figure 2B, and data for PFS are derived from Figure 2A of the published study \protect\cite{Wolchok2022}. For PFS, the subgroup not reaching the 60-month milestone includes 63.82\% of the combined treatment group, 73.67\% of the nivolumab monotherapy group, and 93.81\% of the ipilimumab monotherapy group. For OS, this subgroup includes 47.7\% of the combined treatment group, 56\% of the nivolumab monotherapy group, and 74\% of the ipilimumab monotherapy group.}
\label{fig:comprehensive_analysis_CheckMate067}
\end{figure}

CheckMate 067 provides a representative example of an immunotherapy trial with early overlap followed by later long-term separation. As shown in Figure~\ref{fig:Trial_067_PFSdigitized}, the 6.5-year PFS curves overlap during the first several months and then diverge, with nivolumab plus ipilimumab (Group 2) showing the highest PFS, followed by nivolumab alone (Group 1), and ipilimumab alone (Group 0). Figure~\ref{fig:Trial_067_tau_PFS} shows that the corresponding tau curves begin to rise after this early period, indicating that treatment benefit emerges only after an initial interval of limited separation.

At the 60-month milestone, the estimated long-term progression-free fractions were \(\hat{S}_2^{\text{PFS}}(60)=0.375\), \(\hat{S}_1^{\text{PFS}}(60)=0.301\), and \(\hat{S}_0^{\text{PFS}}(60)=0.062\). The corresponding long-term benefits relative to ipilimumab alone were 0.313 (95\% CI, 0.242--0.384) for the combination regimen and 0.239 (95\% CI, 0.170--0.308) for nivolumab monotherapy, with both \(P\) values close to 0. Among patients who did not reach the 60-month PFS milestone, Figure~\ref{fig:Trial_067_susceptible_tau_PFS} shows that the susceptible tau curve for the combination regimen declines initially and then rises sharply after approximately 3 months, remaining above the corresponding curve for nivolumab monotherapy thereafter. This pattern indicates a clinically relevant trade-off: among patients who ultimately fail by 60 months, the combination regimen shows less favorable early dynamics but stronger subsequent benefit within the short-term subgroup.

For OS at 6.5 years, Figure~\ref{fig:Trial_067_OSdigitized} shows that the three survival curves are similar during the first 3 months and then separate, with ipilimumab alone declining earliest and most sharply. Figure~\ref{fig:Trial_067_tau_OS} shows that the OS tau curves begin to increase after approximately 3 months, again indicating delayed emergence of treatment benefit. At the 60-month milestone, the estimated long-term survival fractions were \(\hat{S}_2^{\text{OS}}(60)=0.523\), \(\hat{S}_1^{\text{OS}}(60)=0.440\), and \(\hat{S}_0^{\text{OS}}(60)=0.260\), corresponding to long-term benefits of 0.263 (95\% CI, 0.188--0.339) for the combination regimen and 0.181 (95\% CI, 0.106--0.256) for nivolumab monotherapy, both relative to ipilimumab alone. However, Figure~\ref{fig:Trial_067_susceptible_tau_OS} shows that among patients who did not reach the 60-month OS milestone, the susceptible tau curve for the combination regimen remains below zero for much of follow-up, whereas the corresponding curve for nivolumab monotherapy remains close to zero. This pattern suggests that an overall long-term survival advantage does not imply uniformly improved short-term outcomes within the subgroup that did not reach the milestone.

\begin{figure}[H]
\centering
\begin{minipage}{.49\textwidth}
    \centering
    \includegraphics[width=\linewidth]{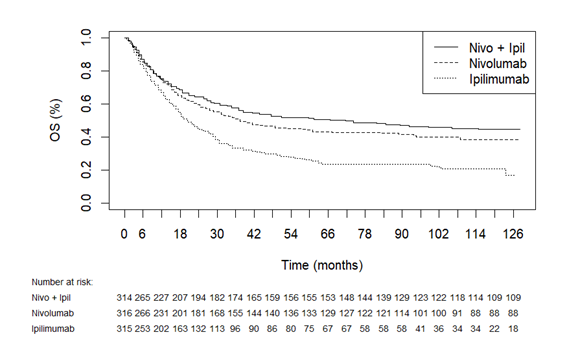}
    \subcaption{Digitized Overall Survival (OS) Curves.}
    \label{fig:Trial_067_final_OS_digitized}
\end{minipage}%
\hfill
\begin{minipage}{.49\textwidth}
    \centering
    \includegraphics[width=\linewidth]{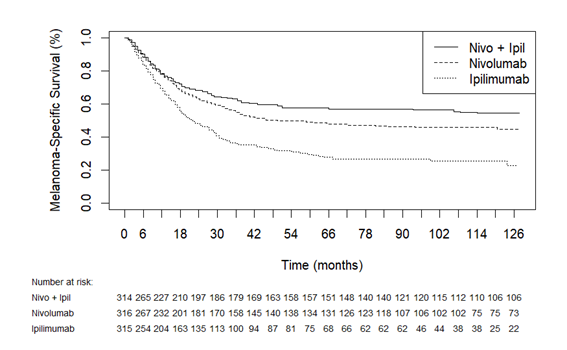}
    \subcaption{Digitized MSS Curves.}
    \label{fig:Trial_067_final_MSS_digitized}
\end{minipage}%

\vspace{1em} 

\begin{minipage}{.49\textwidth}
    \centering
    \includegraphics[width=\linewidth]{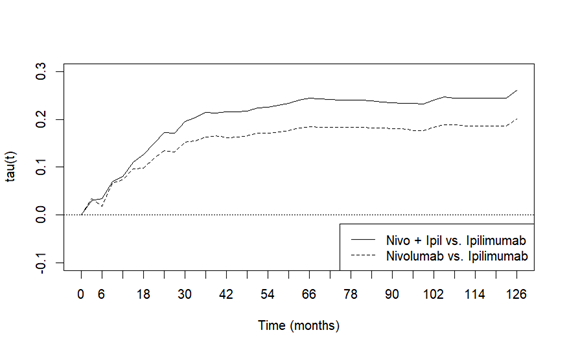}
    \subcaption{Tau Curves for OS.}
    \label{fig:Trial_067_final_tau_OS}
\end{minipage}%
\hfill
\begin{minipage}{.49\textwidth}
    \centering
    \includegraphics[width=\linewidth]{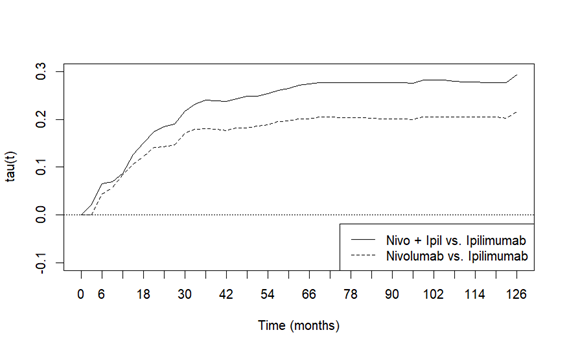}
    \subcaption{Tau Curves for MSS.}
    \label{fig:Trial_067_final_tau_MSS}
\end{minipage}

\vspace{1em} 

\begin{minipage}{.49\textwidth}
    \centering
    \includegraphics[width=\linewidth]{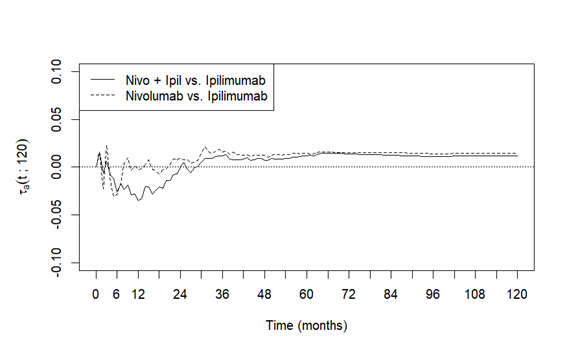}
    \subcaption{Tau Curve for OS in Patients Not Reaching the 120-Month Milestone.}
    \label{fig:Trial_067_final_susceptible_tau_OS}
\end{minipage}%
\hfill
\begin{minipage}{.49\textwidth}
    \centering
    \includegraphics[width=\linewidth]{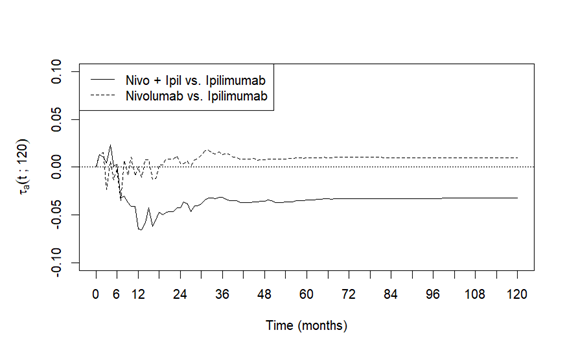}
    \subcaption{Tau Curve for MSS in Patients Not Reaching the 120-Month Milestone.}
    \label{fig:Trial_067_final_susceptible_tau_MSS}
\end{minipage}
\caption{Comprehensive analysis from the final 10-year report of the CheckMate 067 trial, presenting metrics based on OS and melanoma-specific survival (MSS), along with their tau curves. Data for OS and MSS are derived from Figures 1A and 1B of the published study \protect\cite{Wolchok2024}. For OS, the subgroup not reaching the 120-month milestone includes 55.4\% of the combined treatment group, 61.4\% of the nivolumab monotherapy group, and 79.2\% of the ipilimumab monotherapy group. For melanoma-specific survival,  the subgroup not reaching the 120-month milestone includes 45.6\% of the combined treatment group, 54.2\% of the nivolumab monotherapy group, and 74.3\% of the ipilimumab monotherapy group.}
\label{fig:comprehensive_analysis_CheckMate067_final}
\end{figure}

The 10-year analysis further clarifies the distinction between OS and MSS. As shown in Figures~\ref{fig:Trial_067_final_OS_digitized} and \ref{fig:Trial_067_final_MSS_digitized}, both endpoints show durable long-term separation, but MSS exhibits slightly stronger tail separation than OS. This difference is also reflected in the tau curves in Figures~\ref{fig:Trial_067_final_tau_OS} and \ref{fig:Trial_067_final_tau_MSS}, where the MSS curve remains somewhat more favorable over extended follow-up, consistent with the increasing contribution of non--melanoma-related deaths to OS. At the 120-month milestone, the long-term benefit relative to Group 0 was 0.239 (95\% CI, 0.164--0.313) for OS and 0.287 (95\% CI, 0.209--0.366) for MSS in Group 2, compared with 0.179 (95\% CI, 0.105--0.253) for OS and 0.201 (95\% CI, 0.123--0.279) for MSS in Group 1. Figures~\ref{fig:Trial_067_final_susceptible_tau_OS} and \ref{fig:Trial_067_final_susceptible_tau_MSS} further show that, among patients who did not reach the 120-month milestone, the susceptible tau curves again display an early disadvantage for the combination regimen followed by later improvement. Overall, CheckMate 067 illustrates how the proposed framework separates durable long-term survival from early risk within the same trial.

\subsection{CheckMate 227}

\begin{figure}[H]
\centering

\begin{minipage}{.49\textwidth}
    \centering
    \includegraphics[width=\linewidth]{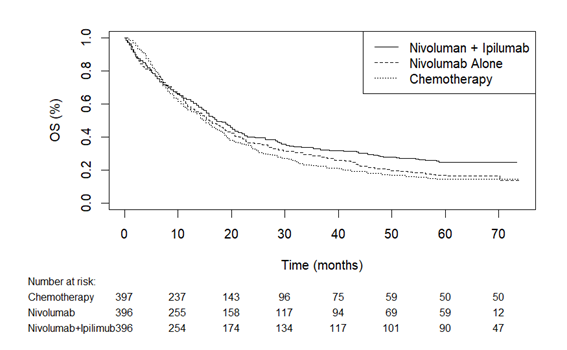} \subcaption{Digitized Overall Survival Curves.}
    \label{fig:trial227osyear5_digitized}
\end{minipage}%
\hfill
\begin{minipage}{.49\textwidth}
    \centering
    \includegraphics[width=\linewidth]{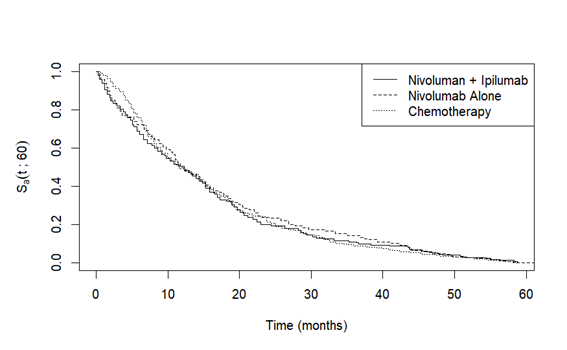}
    \subcaption{Survival Curves in Patients Not reaching the 60-month milestone.}
    \label{fig:susceptibleSurvivalYear5}
\end{minipage}

\vspace{1em} 

\begin{minipage}{.49\textwidth}
    \centering
    \includegraphics[width=\linewidth]{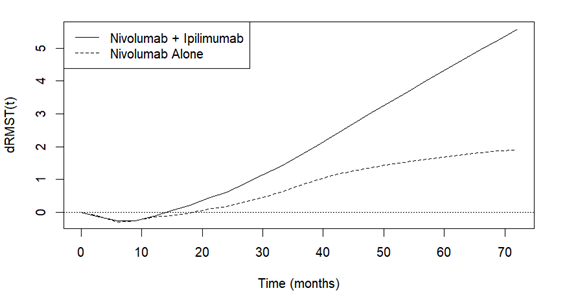}
    \subcaption{dRMST Curves for OS.}
    \label{fig:Trial227_dRMSTYear5}
\end{minipage}%
\hfill
\begin{minipage}{.49\textwidth}
    \centering
    \includegraphics[width=\linewidth]{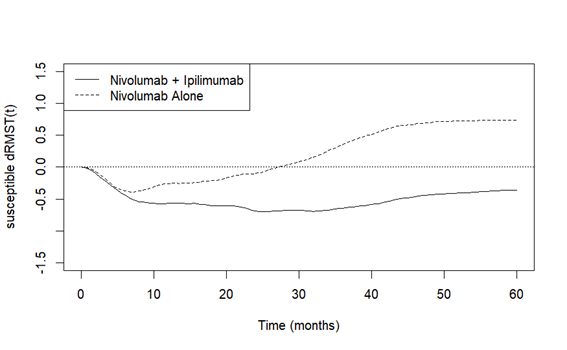}
    \subcaption{dRMST Curves in Patients Not reaching the 60-month milestone.}
    \label{fig:susceptible_dRMSTYear5}
\end{minipage}

\vspace{1em} 

\begin{minipage}{.49\textwidth}
    \centering
    \includegraphics[width=\linewidth]{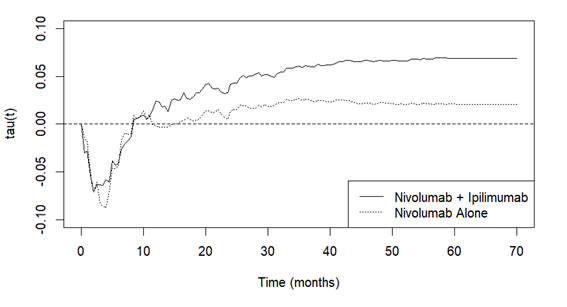}
    \subcaption{Tau Curves for OS.}
    \label{fig:Trial227_tauYear5}
\end{minipage}%
\hfill
\begin{minipage}{.49\textwidth}
    \centering
    \includegraphics[width=\linewidth]{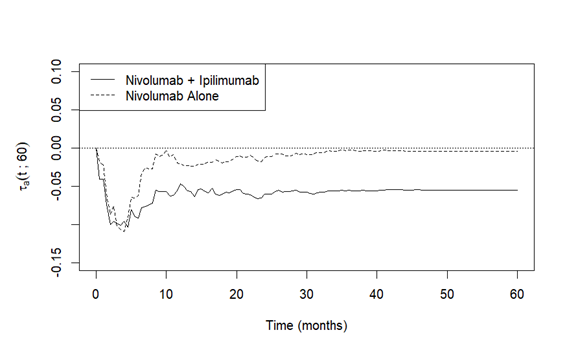}
   \subcaption{Tau curves in Patients Not reaching the 60-month milestone.}
    \label{fig:susceptible_tau5}
\end{minipage}

\caption{Comprehensive analysis from the 5-year CheckMate 227 study, based on digitized OS Kaplan-Meier curves from Figure 2A \protect\cite{Brahmer2023}, for patients with tumor PD-L1 expression levels above 1\%. The subgroup not reaching the 60-month OS milestone includes 75.10\% in the combined treatment group, 83.08\% in nivolumab monotherapy, and 85.51\% in ipilimumab monotherapy.}
\label{fig:comprehensive_analysis_227}
\end{figure}

CheckMate 227 represents a different survival pattern, with early crossing followed by later separation. As shown in Figure~\ref{fig:trial227osyear5_digitized}, the OS curves for nivolumab plus ipilimumab (Group 2), nivolumab alone (Group 1), and chemotherapy (Group 0) are close early in follow-up and then cross at approximately 7 months, indicating that early survival ordering differs from later survival ordering. Thereafter, both nivolumab-containing regimens separate from chemotherapy, with the combination regimen maintaining the strongest long-term plateau. Figure~\ref{fig:Trial227_tauYear5} shows that the corresponding tau curves reach their minima at approximately 3 to 4 months and then turn upward, approaching and subsequently remaining near or above zero by approximately 8 to 10 months. This suggests that improvement in treatment ordering begins before the crossing of the Kaplan--Meier curves becomes visually apparent.

At the 60-month milestone, the long-term survivor fraction for nivolumab plus ipilimumab exceeded that for chemotherapy by approximately 10 percentage points, whereas the corresponding contrast for nivolumab monotherapy was smaller and not statistically significant. Thus, the long-term contrast favors the combination regimen more clearly than nivolumab alone.

Among patients who did not reach the 60-month milestone, Figure~\ref{fig:susceptibleSurvivalYear5} shows that the conditional survival curves crossed within the first year and then exhibited only limited separation thereafter. Figure~\ref{fig:susceptible_tau5} further shows that the susceptible tau curves declined early and remained below zero, with an upward turn around 2 to 4 months and a relatively flat trajectory after approximately 12 months. This pattern indicates that the early disadvantage was concentrated in the first several months and weakened later. In this setting, the milestone framework helps clarify a pattern that is difficult to summarize using a single conventional measure: early unfavorable survival ordering can coexist with later durable benefit, particularly for the combination regimen. Supplementary PH diagnostics and a complementary comparison with the earlier 4-year report are provided in the Supplementary Materials.

\subsection{CLEAR}

\begin{figure}[H]
\centering
\begin{minipage}{.49\textwidth}
    \centering
    \includegraphics[width=\linewidth]{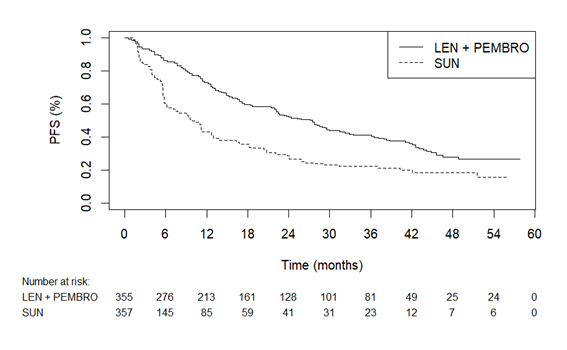}
    \subcaption{Digitized Progression-Free Survival Curves.}
    \label{fig:CLEARtrialPFSdigitized}
\end{minipage}
\hfill
\begin{minipage}{.49\textwidth}
    \centering
    \includegraphics[width=\linewidth]{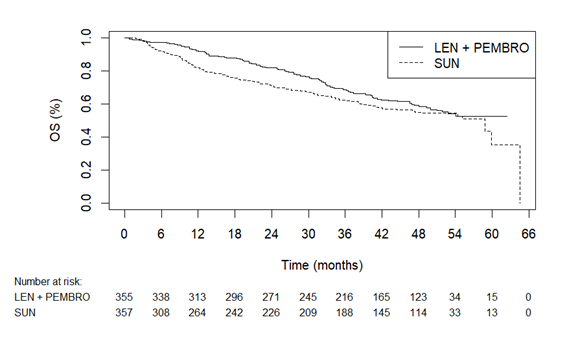}
    \subcaption{Digitized Overall Survival Curves.}
    \label{fig:CLEARtrialOSdigitized}
\end{minipage}%

\vspace{1em} 

\begin{minipage}{.49\textwidth}
    \centering
    \includegraphics[width=\linewidth]{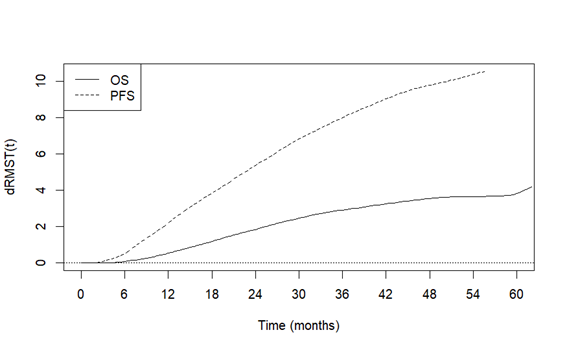}
    \subcaption{dRMST Curves for OS and PFS.}
    \label{fig:CLEARtrial_dRMST_PFS_OS}
\end{minipage}%
\hfill
\begin{minipage}{.49\textwidth}
    \centering
    \includegraphics[width=\linewidth]{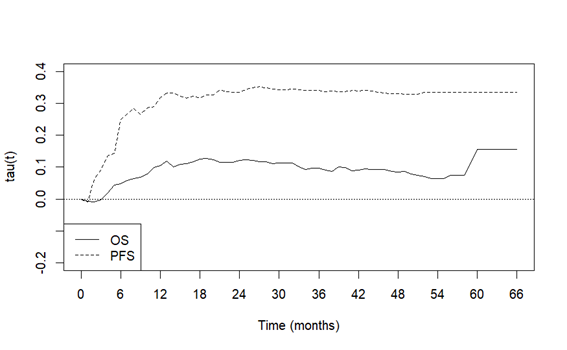}
    \subcaption{Tau Curves for OS and PFS.}
    \label{fig:CLEARtrial_tau_PFS_OS}
\end{minipage}

\vspace{1em} 

\begin{minipage}{.49\textwidth}
    \centering
    \includegraphics[width=\linewidth]{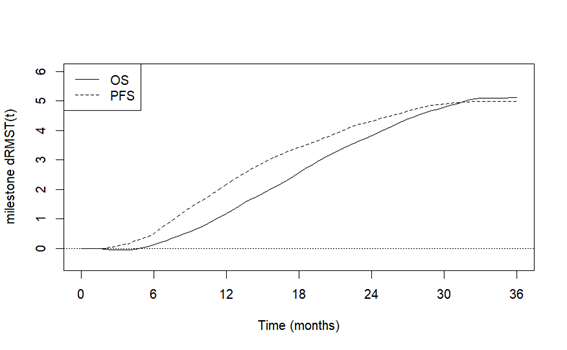}
    \subcaption{dRMST Curves for OS and PFS in Patients Not Reaching the 36-Month Milestone.}
    \label{fig:CLEARtrial_dRMST_susceptible_pfs_os}
\end{minipage}%
\hfill
\begin{minipage}{.49\textwidth}
    \centering
    \includegraphics[width=\linewidth]{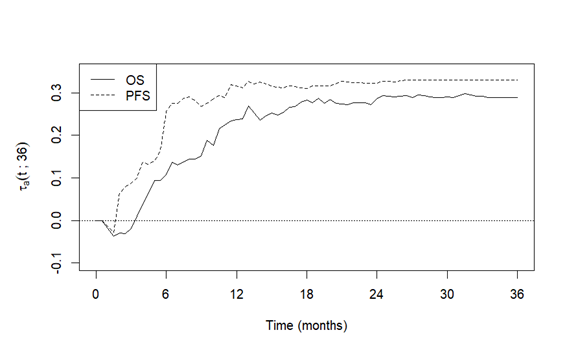}
  \subcaption{Tau Curve for OS and PFS in Patients Not Reaching the 36-Month Milestone.}
   \label{fig:CLEARtrial_tau_susceptible_OS_PFS}
\end{minipage}
\caption{Comprehensive analysis from the CLEAR trial, presenting metrics based on OS and PFS, along with their respective dRMST and tau curves. Data for PFS are derived from Figure 1A and for OS from Figure 2A of the published study \protect\cite{Motzer2024}. In the analysis of PFS, the subgroup not reaching the 36-month milestone includes 58.7\% of the combined treatment group and 77.7\% of the SUN group. For OS, this subgroup includes 31.6\% of the combined treatment group and 37.8\% of the SUN group.}
\label{fig:comprehensive_analysis_CLEAR}
\end{figure}

The CLEAR trial illustrates yet another pattern, with strong and sustained PFS benefit but a more modest and time-limited OS advantage. In this phase III trial, lenvatinib plus pembrolizumab (Group 1) was compared with sunitinib (Group 0) as first-line therapy for advanced renal cell carcinoma. As shown in Figures~\ref{fig:CLEARtrialPFSdigitized} and \ref{fig:CLEARtrialOSdigitized}, the PFS curves show clear and sustained separation over follow-up, whereas the OS curves separate earlier but display greater variability and later convergence. Additional PH diagnostics and remarks on the late-tail behavior of the digitized OS curve are provided in the Supplementary Materials.

The corresponding tau processes further distinguish these two endpoints. Figure~\ref{fig:CLEARtrial_tau_PFS_OS} shows that \(\hat{\tau}^{\text{PFS}}(t)\) rises rapidly early and then stabilizes, indicating that most of the treatment advantage for PFS is established during the first year and changes little thereafter. In contrast, \(\hat{\tau}^{\text{OS}}(t)\) is smaller in magnitude and flattens later, consistent with weaker and less durable separation than for PFS. This contrast is clinically plausible, because PFS more directly reflects disease control under the assigned regimen, whereas OS is additionally influenced by post-progression management and subsequent therapies.

We set \(m=36\) months as the milestone for the CLEAR analysis. For PFS, the estimated long-term survivor fractions were \(\hat{S}_1^{\text{PFS}}(36)=0.413\) and \(\hat{S}_0^{\text{PFS}}(36)=0.223\), corresponding to a long-term benefit of 0.190 (95\% CI, 0.105--0.275; \(P=1.25\times 10^{-5}\)). Among patients who did not reach the 36-month PFS milestone, Figure~\ref{fig:CLEARtrial_tau_susceptible_OS_PFS} shows that the susceptible tau curve rises sharply from approximately 2 to 8 months and remains positive thereafter, with gradual flattening after about 13 months. This pattern indicates that the short-term PFS benefit is established early and then stabilizes.
For OS, the estimated long-term survivor fractions at 36 months were \(\hat{S}_1^{\text{OS}}(36)=0.684\) and \(\hat{S}_0^{\text{OS}}(36)=0.622\), corresponding to a smaller long-term benefit of 0.062 (95\% CI, -0.011 to 0.135; \(P=0.096\)). In contrast, Figure~\ref{fig:CLEARtrial_tau_susceptible_OS_PFS} shows that, among patients who did not reach the 36-month milestone, the susceptible tau curve for OS initially declines and then turns upward from around the third month onward. This indicates early short-term improvement among susceptible patients despite the absence of a statistically significant long-term OS contrast at the 36-month milestone. Taken together, the CLEAR results illustrate how the proposed framework distinguishes a strong and durable benefit for PFS from a weaker and diminishing effect for OS.

\section{Discussion}

The three trial examples illustrate distinct patterns of treatment-effect heterogeneity under nonproportional hazards. CheckMate 067 showed early overlap followed by later separation, whereas CheckMate 227 showed early crossing, likely reflecting the early efficacy of chemotherapy. Because the tau process is based on pairwise ordering of failure times, crossing may weaken the overall signal more than overlap, especially when it occurs after survival probabilities have already declined substantially. Even so, both trials suggest that the combination regimen began to show improved efficacy at approximately 3 months, and milestone stratification did not substantially alter the estimated timing of hazard reversal.
The analyses of CLEAR and CheckMate 067 also suggest that combination therapies may have a stronger effect on progression-free survival (PFS) than on overall survival (OS). Because disease setting, treatment mechanism, and post-progression management differ across trials, the relationship between PFS and OS may not be consistent across settings. \cite{hess2019relationship}
Nevertheless, the tau process provides a useful way to compare different time-to-event endpoints on a common time scale and to examine how treatment effects evolve across endpoints.

In the presence of long-term survivors, the milestone framework separates treatment effects into two clinically meaningful components: the probability of surviving beyond a prespecified milestone and the pattern of outcomes among patients who do not reach that milestone. This provides a more interpretable description of heterogeneous treatment response than a single summary measure alone. Although statistical significance may vary with the choice of milestone, the major turning points in the tau profiles tend to remain relatively stable, suggesting that the overall shape of the tau trajectory may be more informative than isolated significance tests at a single time point.
The framework also provides a practical alternative to literal cure-mixture interpretation. Rather than attempting to identify a truly cured subgroup, which may be difficult under finite follow-up, the milestone approach defines long-term survival relative to clinically meaningful follow-up times. In this way, it preserves the clinical intuition of separating durable benefit from earlier outcomes while avoiding strong identifiability assumptions.

Compared with mean-based summaries such as dRMST, the tau process is less sensitive to tail extremes because its local behavior is naturally down-weighted when survival probabilities become small in both groups. This feature is particularly useful in settings with sparse late risk sets, where visual interpretation of Kaplan--Meier tails may be unstable. More broadly, although we focus here on OS and PFS, the same framework may also be useful for other clinically relevant time-to-event outcomes.
Several limitations should also be noted. First, the analyses are based on reconstructed individual-level data rather than original patient-level trial data, so some approximation error is unavoidable. Second, interpretation of late follow-up remains sensitive to data maturity and to the number of patients remaining at risk. Third, the milestone framework is intended primarily as an interpretive tool and does not eliminate the need for endpoint-specific clinical judgment when selecting milestone times or comparing trial settings.

Overall, the milestone-based framework provides a clinically interpretable framework for oncology trials in which early risk, delayed benefit, and durable long-term survival may coexist. By separating long-term benefit from earlier outcome dynamics and clarifying when treatment effects begin to emerge, it helps reveal treatment-effect patterns that conventional survival summaries may obscure under nonproportional hazards.


\clearpage


\section*{Figures}

\newpage
\bibliographystyle{unsrt}
\bibliography{main_ref} 

\end{document}


\onehalfspacing
\maketitle

\noindent $^{*}$Corresponding authors: Martin T. Wells, mtw1@cornell.edu; Weijing Wang, wjwang@stat.nycu.edu.tw

\section{Supplementary Methods and Supporting Results}

These supplementary materials provide brief methodological details and supporting trial-specific analyses for the main manuscript. To avoid repeating the main-text background and primary interpretations, we focus here on notation, estimation details, remarks on late follow-up, and supplementary analyses that complement the trial summaries reported in the main paper. The trial applications are based on reconstructed individual-level data derived from published Kaplan--Meier curves, as described in the main manuscript.

\subsection{Overall tau process}

Let $j \in \{0,1\}$ index the treatment groups, with $j=0$ denoting the control group and $j=1$ the treatment group. For each group, let $T_j$ be the event time, $S_j(t)=\Pr(T_j>t)$ the survival function, and $F_j(t)=1-S_j(t)$ the cumulative distribution function. The hazard and cumulative hazard functions are denoted by
\[
\lambda_j(t)=\frac{f_j(t)}{S_j(t)}
\qquad \text{and} \qquad
\Lambda_j(t)=\int_0^t \lambda_j(u)\,du,
\]
respectively, where $f_j$ is the density of $T_j$.

The two-sample tau process $\tau(t)$ is a model-free estimand that compares the ordering of event times between the two groups up to time $t$:
\begin{equation}
\tau(t)
= \Pr(T_0 < T_1 \wedge t) - \Pr(T_1 < T_0 \wedge t)
= \int_0^t S_1(u)\,dF_0(u) - \int_0^t S_0(u)\,dF_1(u).
\label{eq:tau_overall}
\end{equation}
The total probability mass contributing to $\tau(t)$ is
\[
1 - \Pr(T_0 \wedge T_1 > t) = 1 - S_0(t)S_1(t),
\]
so $\tau(t)$ can be interpreted as the net probability, among pairs with at least one event by $t$, that the patient in Group~1 survives longer than the patient in Group~0. A positive value of $\tau(t)$ therefore indicates more favorable event-time ordering for Group~1 up to time $t$.

Differentiating \eqref{eq:tau_overall} with respect to $t$ gives the local slope
\begin{equation}
\tau'(t)=S_0(t)S_1(t)\{\lambda_0(t)-\lambda_1(t)\}.
\label{eq:tau_overall_slope}
\end{equation}
The sign of $\tau'(t)$ is determined by the hazard difference $\lambda_0(t)-\lambda_1(t)$, while $S_0(t)S_1(t)$ acts as a weight reflecting the proportion of comparable risk pairs at time $t$. A sign change in $\tau'(t)$ identifies hazard reversal between the two groups.

\subsection{Milestone-based decomposition and susceptible tau process}

Let $m>0$ be a clinically chosen milestone time. For each treatment group $j \in \{0,1\}$, define the long-term survivor fraction
\[
\eta_j(m)=\Pr(T_j>m), \qquad j=0,1.
\]
We also define the susceptible (short-term) survival function
\[
S_{a,j}(t;m)=\Pr(T_j>t\mid T_j\le m),
\qquad 0\le t\le m,\; j=0,1.
\]
Thus, $\eta_j(m)$ represents the long-term component beyond the milestone, whereas $S_{a,j}(t;m)$ describes the subgroup that experiences the event no later than $m$. For $t\le m$, the survival function of group $j$ can be written in mixture form
\begin{equation}
S_j(t)=S_{a,j}(t;m)\{1-\eta_j(m)\}+\eta_j(m),
\qquad j=0,1.
\label{eq:milestone_mixture}
\end{equation}
Within the susceptible region $\{T_j\le m\}$, we define
\[
F_{a,j}(t;m)=1-S_{a,j}(t;m), \qquad j=0,1,
\]
and consider the milestone-stratified susceptible tau process
\begin{equation}
\tau_a(t;m)
= \Pr(T_1 \wedge t > T_0 \mid T_0\le m, T_1\le m)
- \Pr(T_0 \wedge t > T_1 \mid T_0\le m, T_1\le m),
\label{eq:tau_short_def}
\end{equation}
which can be written as
\begin{equation}
\tau_a(t;m)
= \int_0^t S_{a,1}(u;m)\,dF_{a,0}(u;m)
- \int_0^t S_{a,0}(u;m)\,dF_{a,1}(u;m),
\qquad 0\le t\le m.
\label{eq:tau_short}
\end{equation}
A positive value of $\tau_a(t;m)$ indicates that, among patients who do not survive beyond $m$, those in Group~1 tend to have longer event times than those in Group~0 up to time $t$.

Differentiating \eqref{eq:tau_short} yields
\begin{equation}
\tau_a'(t;m)=S_{a,0}(t;m)S_{a,1}(t;m)\{\lambda_{a,0}(t;m)-\lambda_{a,1}(t;m)\},
\label{eq:tau_short_slope}
\end{equation}
where $\lambda_{a,j}(t;m)$ is the hazard function of the conditional distribution $\{T_j\mid T_j\le m\}$. Writing $f_{a,j}(t;m)$ for the corresponding density, we have
\begin{equation}
\lambda_{a,j}(t;m)=\frac{f_{a,j}(t;m)}{\Pr(T_j\in[t,m])},
\qquad t<m,\; j=0,1.
\label{eq:lambda_a}
\end{equation}
Equation~\eqref{eq:tau_short_slope} shows that $\tau_a'(t;m)$ is a weighted hazard difference between the two susceptible risk sets, with weight $S_{a,0}(t;m)S_{a,1}(t;m)$. A sign change in $\tau_a'(t;m)$ therefore indicates hazard reversal within the milestone-defined susceptible subgroup. In this framework, long-term effects are summarized by contrasts such as $\eta_1(m)-\eta_0(m)$, whereas short-term effects are represented by the trajectory of $\tau_a(t;m)$ for $t\le m$.

\subsection{Estimation and inference}

Let $\widehat{S}_j(t)$ be the Kaplan--Meier estimator for group $j$, and let
\[
\widehat{\eta}_j(m)=\widehat{S}_j(m), \qquad j=0,1.
\]
A plug-in estimator for the susceptible survival function is obtained from \eqref{eq:milestone_mixture} as
\begin{equation}
\widehat{S}_{a,j}(t;m)
= \frac{\widehat{S}_j(t)-\widehat{\eta}_j(m)}{1-\widehat{\eta}_j(m)},
\qquad 0\le t\le m,\; j=0,1.
\label{eq:S_a_hat}
\end{equation}
Define
\[
\widehat{F}_{a,j}(t;m)=1-\widehat{S}_{a,j}(t;m),
\]
and obtain the estimated susceptible tau process by replacing $(S_{a,j},F_{a,j})$ with $(\widehat{S}_{a,j},\widehat{F}_{a,j})$ in \eqref{eq:tau_short}:
\[
\widehat{\tau}_a(t;m)
= \int_0^t \widehat{S}_{a,1}(u;m)\,d\widehat{F}_{a,0}(u;m)
- \int_0^t \widehat{S}_{a,0}(u;m)\,d\widehat{F}_{a,1}(u;m),
\qquad 0\le t\le m.
\]
In multi-arm trials with a baseline group $j=0$ and additional groups $j=1,2,\ldots$, we write $\widehat{\tau}_{a,j}(t;m)$ for the susceptible tau process comparing group $j$ with the baseline group. Throughout the trial applications, point estimates, confidence intervals, and $P$ values are obtained by bootstrap resampling based on 10{,}000 resamples.

\subsection{Data maturity and late follow-up}
\label{sec:data_maturity_followup}

Del Paggio and Tannock discuss data maturity as the extent to which follow-up is sufficiently long to support stable interpretation of the primary endpoint, especially in the tail of the survival curve \cite{DelPaggio2021CautionaryTails}. Kaplan--Meier curves are routinely presented together with numbers at risk for this reason \cite{DelPaggio2021CautionaryTails,Hellmann2016MediansMilestones}. Pocock et al. suggested that late-tail interpretation is more reliable when roughly 10\% to 20\% of patients remain at risk \cite{Pocock2002}.

Let $T_i$ be the event time and $C_i$ the censoring time in the original study for $i=1,\ldots,n$. The observed variables are $(X_i,\delta_i)$, where $X_i=T_i\wedge C_i$ and $\delta_i=I(T_i\le C_i)$. The number at risk at time $t$ in the original study is
\[
Y(t)=\sum_{i=1}^n I(X_i\ge t).
\]
When follow-up is extended, the censoring times are updated to $C_i^*$, where $C_i^*\ge C_i$ for each $i$. Define $X_i^*=T_i\wedge C_i^*$ and $\delta_i^*=I(T_i\le C_i^*)$. Since $X_i^*\ge X_i$ for all $i$, it follows that $Y^*(t)\ge Y(t)$ for all $t$, where
\[
Y^*(t)=\sum_{i=1}^n I(X_i^*\ge t).
\]
Under the two follow-up schemes, the corresponding Kaplan--Meier estimators are
\[
\widehat{S}(t)=\prod_{u\le t}\left\{1-\frac{D(u)}{Y(u)}\right\},
\qquad
\widehat{S}^*(t)=\prod_{u\le t}\left\{1-\frac{D^*(u)}{Y^*(u)}\right\},
\]
where $D(u)=\sum_{i=1}^n I(X_i=u,\delta_i=1)$ counts events at time $u$ under the original follow-up, and $D^*(u)=\sum_{i=1}^n I(X_i^*=u,\delta_i^*=1)$ counts events under the extended follow-up. Because $\delta_i\le \delta_i^*$ and $Y(u)\le Y^*(u)$, there is no simple monotone ordering between $\widehat{S}(t)$ and $\widehat{S}^*(t)$.

Nevertheless, the tail behavior of these estimators remains informative. Let $t_{\max}$ denote the maximum failure time under the original follow-up scheme. It is possible that $\widehat{S}(t_{\max})=0$, which occurs when $D(t_{\max})=Y(t_{\max})$, whereas $\widehat{S}^*(t_{\max})>0$ if $Y^*(t_{\max})>D^*(t_{\max})$. When $\widehat{S}(t_{\max})=0$, it is also useful to examine the smallest positive jump in the estimator near $t_{\max}$. If that value remains clearly above zero but is based on a very small number at risk, the apparent drop to zero may not be reliable. This issue arises in the CLEAR OS analysis, where the sunitinib Kaplan--Meier curve drops to zero near the end of follow-up when only a small number of patients remain at risk.

\section{CheckMate 067: Additional Analyses}
\label{sec:supp067}

The main manuscript presents the primary CheckMate~067 results from the 6.5-year PFS and OS analyses and from the final 10-year OS and MSS update. The sections below provide supporting diagnostics and numerical summaries for the trial applications.

\subsection{Supporting analyses for the 6.5-year digitized data (PFS and OS)}

\begin{figure}[htbp]
\centering
\includegraphics[width=\textwidth]{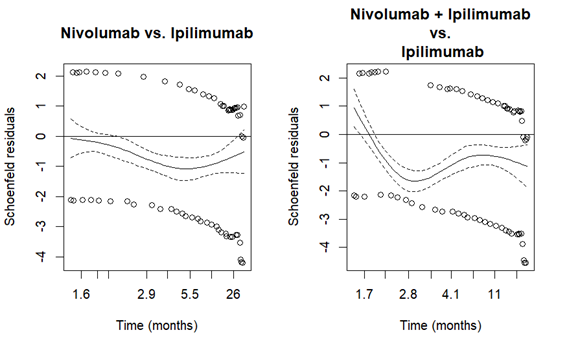}
\caption{Schoenfeld residual plots for PFS based on the 6.5-year digitized data from CheckMate~067 \cite{Wolchok2022}. The $P$ values obtained from testing the PH assumption are 0.004 for the comparison between nivolumab alone and ipilimumab alone (left), and 0.056 for the comparison between nivolumab plus ipilimumab and ipilimumab alone (right).}
\label{fig:checkPH_PFS_067}
\end{figure}

\begin{figure}[htbp]
\centering
\includegraphics[width=0.85\textwidth]{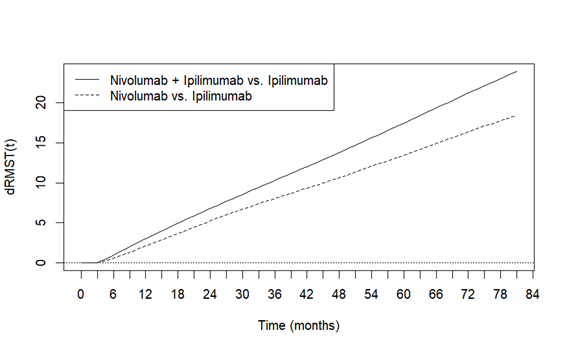}
\caption{Estimated dRMST curves for PFS from the digitized CheckMate~067 data (Fig.~2A in \citealp{Wolchok2022}). Both curves exhibit a linear and steady upward trend, beginning from approximately the third month onward. For Group~2 versus Group~0, $\widehat{\mu}_2^D(81.13)=23.943$ with a 95\% confidence interval of $(18.930,\ 28.956)$; for Group~1 versus Group~0, $\widehat{\mu}_1^D(81.13)=18.476$ with a 95\% confidence interval of $(13.575,\ 23.378)$. Both $P$ values are close to zero, indicating strong statistical significance. Note that the dRMST curves are influenced by tail areas where the survival curves approach plateaus, as indicated in Fig.~2A of \citet{Wolchok2022}.}
\label{fig:dRMST_PFS_067}
\end{figure}

\begin{figure}[htbp]
\centering
\includegraphics[width=0.85\textwidth]{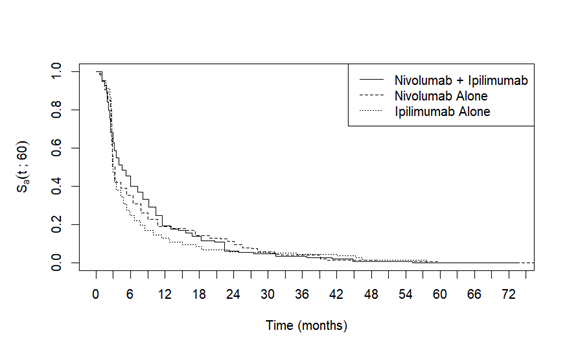}
\caption{Progression-free survival curves for the subgroup of patients from the CheckMate~067 trial \cite{Wolchok2022} who did not reach the 60-month milestone. The curves indicate that a substantial fraction of these patients experienced the event in the first 3 months. In the middle portion of follow-up, roughly $t\in[3,12]$ months, Group~2 appears more favorable than Group~1, whereas Group~0 is less favorable from approximately $t=3$ months onward.}
\label{fig:Trial_067_Susceptible_PFS}
\end{figure}

\begin{figure}[htbp]
\centering
\includegraphics[width=0.85\textwidth]{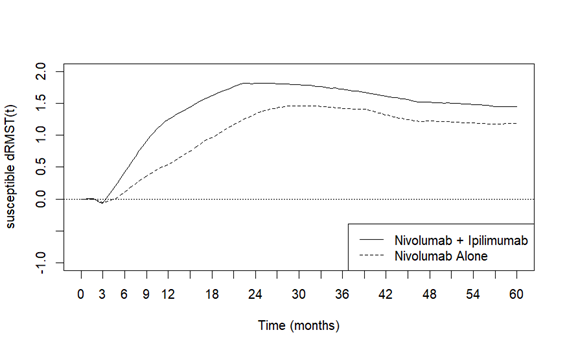}
\caption{PFS dRMST curves for the subgroup of patients from the CheckMate~067 trial \cite{Wolchok2022} who did not reach the 60-month milestone. The two curves begin to diverge at approximately the third month. For Group~1 versus Group~0, the dRMST curve becomes positive from about $t\approx 3$ months onward with $\widehat{\mu}^{D}_{a,1}(60;m=60)=1.186$, 95\% CI $(-1.277,\ 3.650)$, and $P=0.345$. For Group~2 versus Group~0, the dRMST is positive and shows an increasing pattern from about $t\approx 3$ months with $\widehat{\mu}^{D}_{a,2}(60;m=60)=1.446$, 95\% CI $(-0.965,\ 3.856)$, and $P=0.240$. These results suggest modest differences in short-term PFS after conditioning on not reaching the 60-month milestone, but neither comparison is statistically significant.}
\label{fig:Trial_067_Susceptible_dRMST_PFS}
\end{figure}

For progression-free survival, Figure~\ref{fig:checkPH_PFS_067} presents PH diagnostics for Group~2 versus Group~0 and Group~1 versus Group~0, and Figure~\ref{fig:dRMST_PFS_067} shows the corresponding dRMST processes. For completeness, the overall $\tau$-process estimates (Group~$j$ versus Group~0) at the end of follow-up are $\widehat{\tau}^{\mathrm{PFS}}_{2}(81.3)=0.3807$ (95\% CI, $0.291$ to $0.471$) and $\widehat{\tau}^{\mathrm{PFS}}_{1}(81.3)=0.2578$ (95\% CI, $0.164$ to $0.351$), with both two-sided $P$ values $<0.001$. At the end of follow-up, the estimated PFS probabilities were 0.3618 (Group~2, $t=73.5$ months), 0.2633 (Group~1, $t=80.08$ months), and 0.0619 (Group~0, $t=57.61$ months).
At the milestone $m=60$ months, Figure~\ref{fig:Trial_067_Susceptible_PFS} reports the conditional PFS curves $\widehat{S}^{\mathrm{PFS}}_{a,j}(t;60)$ for $j=0,1,2$ in the subgroup that does not reach the milestone, and Figure~\ref{fig:Trial_067_Susceptible_dRMST_PFS} shows the associated milestone-stratified dRMST contrasts $\widehat{\mu}^{D}_{a,j}(t;60)$ for $j=1,2$ versus Group~0. The corresponding subgroup proportions (not reaching $m=60$) are 63.82\% in Group~2, 73.67\% in Group~1, and 93.81\% in Group~0. The formal susceptible $\tau$ contrasts at $t=60$ months are $\widehat{\tau}^{\mathrm{PFS}}_{a,2}(60;60)=0.117$ (95\% CI, $-0.004$ to $0.238$; $P=0.058$) and $\widehat{\tau}^{\mathrm{PFS}}_{a,1}(60;60)=0.032$ (95\% CI, $-0.081$ to $0.146$; $P=0.579$).

\begin{figure}[htbp]
\centering
\includegraphics[width=\textwidth]{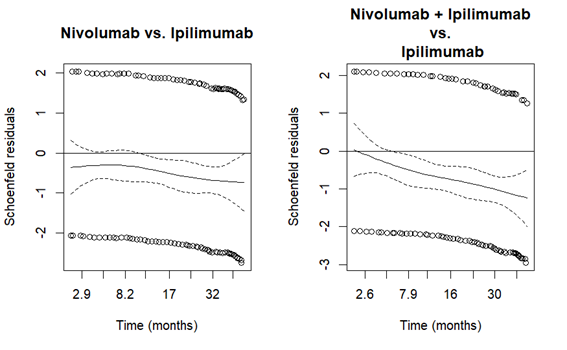}
\caption{Schoenfeld residual plots for OS based on the 6.5-year digitized data from CheckMate~067 \cite{Wolchok2022}. The $P$ values obtained from testing the PH assumption are 0.14 for the comparison between nivolumab alone and ipilimumab alone (left), indicating no significant violation of the PH assumption, and 0.0011 for the comparison between nivolumab plus ipilimumab combination therapy and ipilimumab alone (right), indicating evidence against proportional hazards for the combination-therapy comparison.}
\label{fig:checkPH_OS_067}
\end{figure}

\begin{figure}[htbp]
\centering
\includegraphics[width=0.85\textwidth]{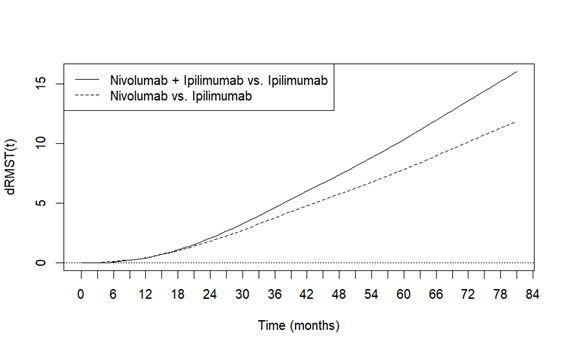}
\caption{Estimated dRMST curves for OS from the digitized CheckMate~067 data (Fig.~2B in \citealp{Wolchok2022}). Both curves exhibit a steady upward trend, showing positive gains beginning around the sixth month and further divergence beginning around the 18th month. The combination therapy (Group~2) appears more favorable than monotherapy. For Group~2 versus Group~0, the estimated $\widehat{\mu}_2^D(85.29)=17.232$ with a 95\% confidence interval of $(11.866,\ 22.598)$; for Group~1 versus Group~0, the estimated $\widehat{\mu}_1^D(84.88)=12.627$ with a 95\% confidence interval of $(7.342,\ 17.912)$. Both $P$ values are close to zero, demonstrating statistical significance. Note that the dRMST curves are influenced by tail areas where the survival curves approach plateaus, as indicated in Fig.~2B of \citet{Wolchok2022}.}
\label{fig:dRMST_OS_067}
\end{figure}

\begin{figure}[htbp]
\centering
\includegraphics[width=0.85\textwidth]{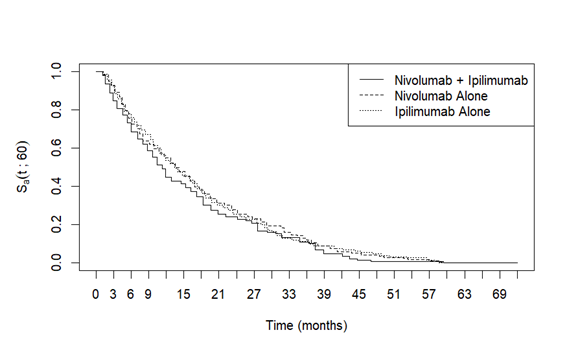}
\caption{Overall survival curves for the subgroup of patients from the CheckMate~067 trial \cite{Wolchok2022} who did not reach the 60-month milestone. The survival function $\widehat{S}_{a,2}^{\mathrm{OS}}(t)$ remains below the other two group-specific curves over most of the observed period.}
\label{fig:Trial_067_Susceptible_OS}
\end{figure}

\begin{figure}[htbp]
\centering
\includegraphics[width=0.85\textwidth]{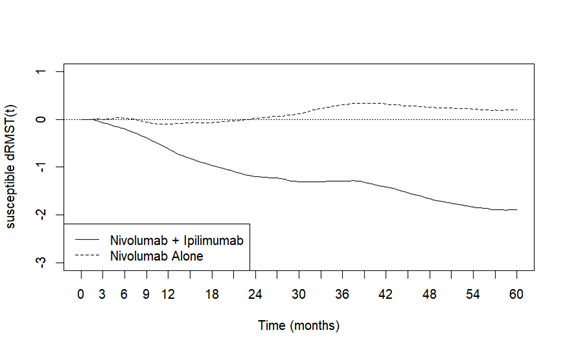}
\caption{OS dRMST curves for the subgroup of patients from the CheckMate~067 trial \cite{Wolchok2022} who did not reach the 60-month milestone. For Group~2 versus Group~0, the dRMST is negative and shows a decreasing pattern with $\widehat{\mu}_{a,2}^D(60;m=60)=-1.888$, 95\% CI $(-4.649,\ 0.874)$, and $P=0.18$. For Group~1 versus Group~0, the dRMST curve turns slightly positive after $t=20$ months, with $\widehat{\mu}_{a,1}^D(60;m=60)=0.198$, 95\% CI $(-2.572,\ 2.967)$, and $P=0.889$.}
\label{fig:Trial_067_Susceptible_dRMST_OS}
\end{figure}

For overall survival, Figure~\ref{fig:checkPH_OS_067} presents PH diagnostics for Group~2 versus Group~0 and Group~1 versus Group~0, and Figure~\ref{fig:dRMST_OS_067} shows the corresponding dRMST processes. For completeness, the overall $\tau$-process estimates (Group~$j$ versus Group~0) at the end of follow-up are $\widehat{\tau}^{\mathrm{OS}}_{2}(85.29)=0.2410$ (95\% CI, $0.153$ to $0.329$) and $\widehat{\tau}^{\mathrm{OS}}_{1}(84.88)=0.1925$ (95\% CI, $0.104$ to $0.281$), with both two-sided $P$ values $<0.001$.
At the milestone $m=60$ months, Figure~\ref{fig:Trial_067_Susceptible_OS} reports the conditional OS curves $\widehat{S}^{\mathrm{OS}}_{a,j}(t;60)$ for $j=0,1,2$ in the subgroup that does not reach the milestone, and Figure~\ref{fig:Trial_067_Susceptible_dRMST_OS} shows the associated milestone-stratified dRMST contrasts. The corresponding subgroup proportions (not reaching $m=60$) are 47.7\% in Group~2, 56.0\% in Group~1, and 74.0\% in Group~0. The formal susceptible $\tau$ contrasts at $t=60$ months are $\widehat{\tau}^{\mathrm{OS}}_{a,2}(60;60)=-0.089$ (95\% CI, $-0.107$ to $0.126$; $P=0.876$) and $\widehat{\tau}^{\mathrm{OS}}_{a,1}(60;60)=0.009$ (95\% CI, $-0.210$ to $0.032$; $P=0.151$).

\subsection{Ten-year milestone contrasts used in the main text (OS and MSS)}

For completeness, we report the 10-year overall survival (OS) and melanoma-specific survival (MSS) milestone contrasts underlying Figure~2 of the main manuscript, based on the final 10-year report from CheckMate~067 \cite{Wolchok2025}. With extended follow-up, more long-term survivors may die from causes unrelated to melanoma; MSS therefore provides a disease-focused endpoint that is less influenced by competing risks.

At the 120-month milestone, the estimated OS probabilities are $\widehat{S}_2^{\mathrm{OS}}(120)=0.446$, $\widehat{S}_1^{\mathrm{OS}}(120)=0.386$, and $\widehat{S}_0^{\mathrm{OS}}(120)=0.208$ for Groups~2, 1, and 0, respectively. The corresponding long-term contrasts relative to Group~0 are $0.239$ (95\% CI, $0.164$ to $0.313$) for Group~2 and $0.179$ (95\% CI, $0.105$ to $0.253$) for Group~1, with both two-sided $P$ values $<0.001$. For MSS at $m=120$, the estimated survival probabilities are $\widehat{S}_2^{\mathrm{MSS}}(120)=0.544$, $\widehat{S}_1^{\mathrm{MSS}}(120)=0.458$, and $\widehat{S}_0^{\mathrm{MSS}}(120)=0.257$. The long-term contrasts relative to Group~0 are $0.287$ (95\% CI, $0.209$ to $0.366$) for Group~2 and $0.201$ (95\% CI, $0.123$ to $0.279$) for Group~1, again with both two-sided $P$ values $<0.001$. Overall, the MSS contrasts are slightly larger than the OS contrasts at 120 months.

\clearpage

\clearpage
\section{CheckMate 227: Additional Analyses}
\label{sec:supp227}

The main manuscript presents the primary CheckMate~227 application using the 5-year OS analysis. The supplementary material here adds supporting diagnostics and a brief comparison with the earlier 4-year report, while keeping the emphasis on the 5-year analysis used in the main text.

\subsection{Five-year analysis (milestone $m=60$ months)}

CheckMate~227 is a phase III trial in advanced NSCLC comparing nivolumab plus ipilimumab (Group~2), nivolumab alone (Group~1), and platinum-based chemotherapy (Group~0) among patients with tumor PD-L1 expression $\ge 1\%$ \cite{Hellmann2019}. Reported minimum follow-up durations include 29.3 months \cite{Hellmann2019}, 4 years \cite{PazAres2022}, and 61.3 months \cite{Brahmer2023}. The primary application in the main manuscript is based on the 5-year report by \citet{Brahmer2023}, using reconstructed individual-level data derived from the published Kaplan--Meier curves via the algorithm of \citet{Guyot2012}. The digitized OS curves, milestone-stratified susceptible curves, and the corresponding dRMST and $\tau$ processes are shown in Figure~3 of the main manuscript.

\begin{figure}[htbp]
\centering
\includegraphics[width=\textwidth]{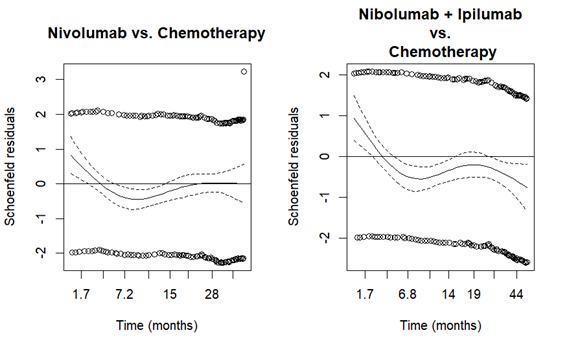}
\caption{Schoenfeld residual plots derived from the 5-year digitized data of CheckMate~227 \cite{Brahmer2023}. The $P$ values obtained from testing the PH assumption are 0.38 for the comparison between nivolumab alone and chemotherapy (left) and 0.002 for the comparison between nivolumab plus ipilimumab and chemotherapy (right), indicating evidence against proportional hazards for the combination-therapy comparison.}
\label{fig:trial227checkPHYear5}
\end{figure}

Figure~\ref{fig:trial227checkPHYear5} reports Schoenfeld residual diagnostics for the PH assumption (Group~2 versus Group~0 and Group~1 versus Group~0). The evidence against PH is stronger for the combination regimen versus chemotherapy ($P=0.002$) than for nivolumab alone versus chemotherapy ($P=0.38$), supporting time-varying treatment effects. To complement the main-text summaries, we also note the turning time of the $\tau$ curve, defined here as the time at which the trajectory shifts from decreasing to increasing. In Figure~3e of the main manuscript, the turning time occurs earlier for the combination regimen (minimum at approximately 2.5 months) than for nivolumab monotherapy (minimum at approximately 4.5 months), suggesting an earlier upward turn for the combination regimen.

\subsection{Supplementary comparison with the earlier 4-year report (milestone $m=48$ months)}
\label{sec:fouryear}

For completeness, we also summarize the milestone survivor proportions from the earlier 4-year report and the later 5-year report. These results are included to provide supplementary context rather than to introduce a separate primary analysis.
At $m=48$ months \citep{PazAres2022}, the estimated OS milestone probabilities are $\widehat{\eta}^{\mathrm{OS}}_{0}(48)=0.180$, $\widehat{\eta}^{\mathrm{OS}}_{1}(48)=0.207$, and $\widehat{\eta}^{\mathrm{OS}}_{2}(48)=0.291$. At $m=60$ months \citep{Brahmer2023}, the long-term survivor contrast is $\widehat{\eta}^{\mathrm{OS}}_{2}(60)-\widehat{\eta}^{\mathrm{OS}}_{0}(60)=0.104$ (95\% CI, $0.047$ to $0.161$; $P=0.00016$), whereas $\widehat{\eta}^{\mathrm{OS}}_{1}(60)-\widehat{\eta}^{\mathrm{OS}}_{0}(60)=0.024$ (95\% CI, $-0.028$ to $0.077$; $P=0.183$), indicating a higher long-term survivor fraction for the combination regimen but not for nivolumab alone.

Within the susceptible subgroup (OS $\le 60$ months; Figure~3b and Figure~3f in the main manuscript), the susceptible $\tau$ curves decline early and remain below zero, but they turn upward within the first few months (around $t\approx 2$--$4.5$ months) and then become nearly flat after about 12 months. This pattern suggests that, within the milestone-defined susceptible subgroup, most of the survival ordering difference occurs during the first few months, with little further change later in follow-up.

\clearpage

\clearpage
\section{CLEAR: Additional Analyses}
\label{sec:suppclear}

\begin{figure}[htbp]
\centering
\includegraphics[width=\textwidth]{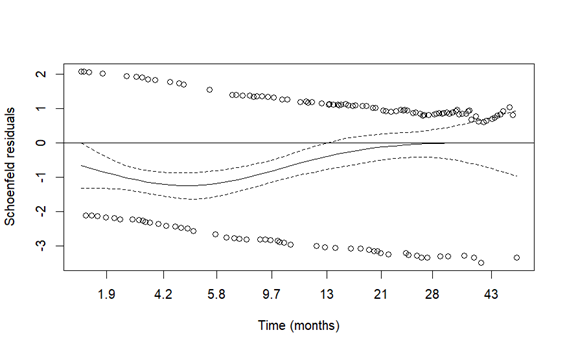}
\caption{Schoenfeld residual plots for progression-free survival (PFS) based on digitized data from the CLEAR trial \cite{Motzer2024}. The PH test $P$ value is 0.00024 for the comparison between Group~1 (lenvatinib plus pembrolizumab) and Group~0 (sunitinib), indicating evidence against proportional hazards.}
\label{fig:checkPH_PFS_CLEAR}
\end{figure}

\begin{figure}[htbp]
\centering
\includegraphics[width=\textwidth]{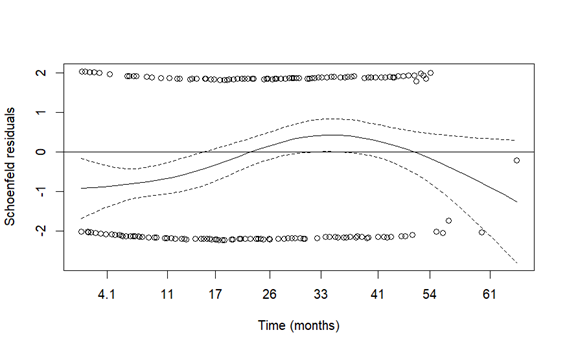}
\caption{Schoenfeld residual plots for overall survival (OS) based on digitized data from the CLEAR trial \cite{Motzer2024}. The PH test $P$ value is 0.002 for the comparison between Group~1 (lenvatinib plus pembrolizumab) and Group~0 (sunitinib), indicating evidence against proportional hazards.}
\label{fig:checkPH_OS_CLEAR}
\end{figure}

The main manuscript emphasizes the contrast between PFS and OS in CLEAR within the milestone-based framework. The additional material collected here focuses on PH diagnostics and on the interpretation of the late tail of the digitized OS curve, where the risk sets become small.
We analyzed digitized Kaplan--Meier curves from the final prespecified overall survival report of the CLEAR phase III trial in advanced renal cell carcinoma \cite{Motzer2024}. Group~1 corresponds to lenvatinib plus pembrolizumab, and Group~0 corresponds to sunitinib. Figures~\ref{fig:checkPH_PFS_CLEAR} and \ref{fig:checkPH_OS_CLEAR} show Schoenfeld residual plots for PFS and OS based on the digitized data. The corresponding tests indicate departures from the PH assumption for both endpoints.

\subsection{Late follow-up behavior of the digitized OS curve}

In the digitized OS curves shown in Figure~4b of the main manuscript, the two groups separate early but later converge and intersect at approximately $t\approx 54$ months, when the risk sets become small. Near the end of follow-up, the sunitinib Kaplan--Meier estimate drops to 0 at the last observed event time (around $t\approx 66$ months), because only a few patients remained at risk and those patients experienced events near the end of follow-up. Consequently, the Kaplan--Meier tail can be highly sensitive to one or a few late events, and OS patterns in the far tail (beyond roughly 54 months) should be interpreted cautiously.

\subsection{Milestone-based results at $m=36$ months}

We set $m=36$ months as the milestone. For PFS, the estimated milestone survival probabilities are $\widehat{S}^{\mathrm{PFS}}_1(36)=0.413$ and $\widehat{S}^{\mathrm{PFS}}_0(36)=0.223$, yielding a milestone contrast of 0.190 (95\% CI, 0.105 to 0.275; $P=1.25\times 10^{-5}$). Among patients who do not reach the 36-month milestone, the susceptible curve $\widehat{\tau}^{\mathrm{PFS}}_a(t;36)$ increases sharply during approximately 2--8 months, continues to increase up to about 13 months, and then stabilizes through $t=36$ months; the 95\% CI for $\widehat{\tau}^{\mathrm{PFS}}_a(36;36)$ is $(0.216,\ 0.444)$.

For OS, the estimated milestone survival probabilities are $\widehat{S}^{\mathrm{OS}}_1(36)=0.684$ and $\widehat{S}^{\mathrm{OS}}_0(36)=0.622$, yielding a smaller milestone contrast of 0.062 (95\% CI, $-0.011$ to 0.135; $P=0.096$). Within the susceptible subgroup (OS $\le 36$ months), the curve $\widehat{\tau}^{\mathrm{OS}}_a(t;36)$ decreases early but increases from approximately the third month onward, with the upward trend becoming less pronounced after about a year; at $t=36$, the 95\% CI for $\widehat{\tau}^{\mathrm{OS}}_a(36;36)$ is $(0.146,\ 0.432)$. At $m=36$ months, the milestone contrast is greater for PFS than for OS, and the late OS estimates are based on smaller risk sets.

\clearpage

\clearpage
\bibliographystyle{unsrtnat}
\bibliography{tail_refs}